\documentclass{aa}                   
\usepackage{graphicx}                

\newcommand{\kms}{km\,s$^{-1}$}

\begin{document}

\title{First time-series optical photometry from Antarctica}
\subtitle{sIRAIT monitoring of the RS~CVn binary V841~Centauri and
the $\delta$-Scuti star V1034~Centauri}

\author{K. G. Strassmeier\inst{1}, R.~Briguglio\inst{2,3},
T.~Granzer\inst{1}, G.~Tosti\inst{2}, I.~DiVarano\inst{1},
I.~Savanov\inst{1}, M.~Bagaglia\inst{2}, S.~Castellini\inst{2},
A.~Mancini\inst{2}, G.~Nucciarelli\inst{2}, O.~Straniero\inst{4},
E.~Distefano\inst{5}, S.~Messina\inst{5} \and G.~Cutispoto\inst{5} }

\offprints{K. G. Strassmeier}

\institute{Astrophysical Institute Potsdam (AIP), An der
Sternwarte 16, D-14482 Potsdam, Germany;\\
\email{kstrassmeier@aip.de}, \email{tgranzer@aip.de},
\email{idivarano@aip.de}, \email{isavanov@rambler.ru} \and
Dipartimento di Fisica, Universit\`a
di Perugia, Via A. Pascoli, I-06100 Perugia, Italy;\\
\email{runa@fisica.unipg.it}, \email{gino.tosti@fisica.unipg.it}
\and Concordia, Dome C, Antarctica; http://www.concordiabase.eu \and
INAF Osservatorio Teramo, Via Mentore Maggini, I-64100 Teramo, Italy
\and INAF - Catania Astrophysical Observatory, via S. Sofia 78,
I-95123 Catania, Italy;\\
\email{sme@oact.inaf.it}, \email{gcutispoto@oact.inaf.it},
\email{edistefano@oact.inaf.it}}

\date{Received ... ; accepted ...}

\abstract{Beating the Earth's day-night cycle is mandatory for long
and continuous time-series photometry and had been achieved with
either large ground-based networks of observatories at different
geographic longitudes or when conducted from space. A third
possibility is offered by a polar location with
astronomically-qualified site characteristics.} {In this paper, we
present the first scientific stellar time-series optical photometry
from Dome~C in Antarctica and analyze approximately 13,000 CCD
frames taken in July 2007.} {The optical pilot telescope of the
``International Robotic Antarctic Infrared Telescope'', named
``small IRAIT'' (sIRAIT), and its $UBVRI$ CCD photometer were used
in $BVR$ for a continuous 243 hours (10.15 days) with a duty cycle
of 98\% and a cadence of 155~sec. The prime targets were the
chromospherically active, spotted binary star V841~Cen and the
non-radially pulsating $\delta$-Scuti star V1034~Cen.}{We confirm
the known 0.2-day fundamental period of V1034~Cen and found a total
of 23 further periods between 2.2 hours and 3.5 days. V841~Cen's $V$
amplitude due to spots appeared to be at a record high of 0\fm4 in
$V$ in July 2007. We present a spot-model analysis with a
light-curve inversion technique and found the star with a spot
filling factor of 44\% of the visible hemisphere, among the highest
ever measured values for active stars, and a temperature difference
photosphere minus spot of 750$\pm$100~K. Its odd-numbered (for a
single site) rotation period was determined with a higher precision
than before (5.8854$\pm$0.0026 days) despite the comparably short
data set. The rms scatter from a 2.4-hour data subset was 3~mmag in
$V$ and 4.2~mmag in $R$. The differential data quality is 3--4 times
better than with the 25cm Fairborn Automatic Photoelectric Telescope
in southern Arizona and is likely due to the exceptionally low
scintillation noise at Dome~C.}{We conclude that high-precision CCD
photometry with exceptional time coverage and cadence can be
obtained at Dome~C in Antarctica and be successfully used for
time-series astrophysics.}

\authorrunning{K. G. Strassmeier et al.}
\titlerunning{Time-series optical photometry from Antarctica}

\maketitle


\section{Introduction}

Time-series photometry is a powerful tool to understand cosmic
variabilities and their many underlying physical mechanisms, from
Gamma Ray Bursts to the non-radial oscillations of our Sun.
World-wide networks around the globe were organized to bypass the
Earth's day-night cycle, e.g. by the ``Global Oscillation Network
Group'' (Harvey et al. \cite{gong}), the ``Whole Earth Telescope''
(Nather et al. \cite{wet}) or the ``Multi Site Continuous
Spectroscopy'' campaigns (Catala et al. \cite{musicos}), just to
name a few. Despite their tremendous success such networks must
cope with the many different site characteristics, the individual
weather patterns and, most importantly, with the different
instrument/detector combinations and the respective calibration
issues. An alternative to such networks is a single polar site.

The French-Italian Antarctic station \emph{Concordia} at Dome~C at
a height of 3200m on the east Antarctic plateau is currently,
besides the U.S. South Pole station, the only plateau station
populated over winter (=night). In principle, it thus enables
astronomical observations comparable to regular observatories at
temperate sites. Dome~C had received world-wide attention from the
night-time astronomical community when it became known that the
seeing conditions on the east Antarctic plateau are likely the
best on the entire planet with a median seeing of 0.3\arcsec\ and
occasionally even below 0.1\arcsec\ at a height of approximately
30m above ground (Lawrence et al. \cite{law:nature}, Agabi et al.
\cite{agabi06}).

\begin{figure*}[ht]
    \centering
    \begin{minipage}[t]{12cm}
       \vspace{0pt}
       \centering
    \includegraphics[width=11cm]{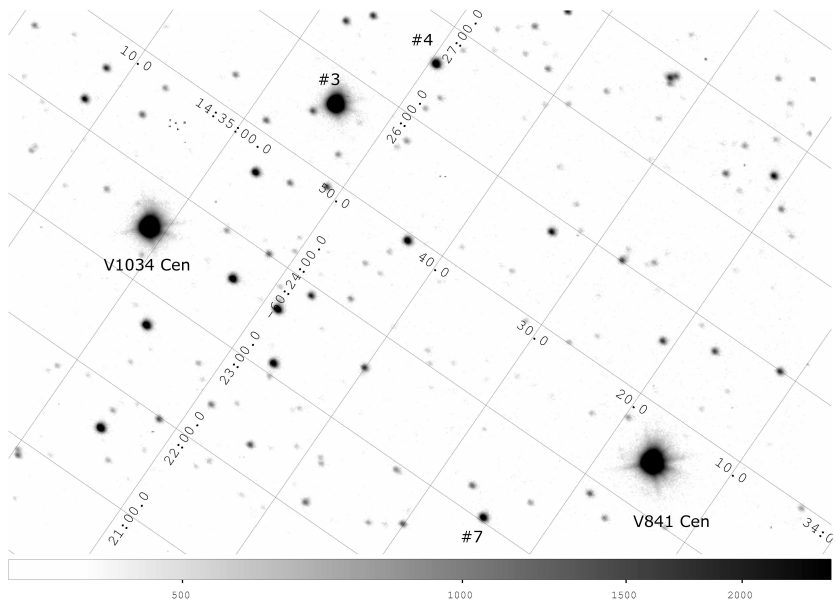}
    \end{minipage}%
    \begin{minipage}[t]{5.8cm}
       \vspace{0pt}
       \centering
    \caption{Identification of stars in the CCD FOV (8$\arcmin\times$5.3\arcmin ).
The three bright stars are our primary targets V841~Cen (right,
$V$=8\fm5) and V1034~Cen (left) and the comparison star CD-59\degr
5309 (top). The comparison star and stars \#4 and \#7 were
identified as variable (see Table~\ref{T1}). The image shown is a
composite of 20 individual R-band frames, each a 40-sec integration.
The coordinates are for equinox 2000.0. The faintest stars are 16th
magnitude.}
    \label{F1}
    \end{minipage}
\end{figure*}

Science cases that require continuous high-precision photometry
uniquely benefit from the absent 24-hour day-night cycle and the
consequently stable atmosphere in general. For example,
$\delta$~Scuti stars are known to have a complex surface oscillation
spectrum involving many modes and frequencies. The detection of
patterns of closely-spaced peaks in different modes enables the
determination of the internal stellar structure. The essential
observational restriction is the frequency resolution, which is
proportional to the length of the photometric time series. It sets
limits to the mode identification and thus their unique
interpretation. Another example are spotted stars. Magnetic spots,
like those on the Sun, are tracers of the internal dynamo activity.
Spot size and temperature are related to the magnetic flux that the
emerging flux tube transported up from the interior. However,
surface velocity fields like differential rotation affect the
magnetic field (and vice versa) so that spots are likely only
indirectly linked to the dynamo. Observing the migration behavior of
starspots from one stellar rotation to the other, however, may
constrains some global properties of the dynamo.

Additionally, optical photometry of bright stars may be brought down
to the photon-noise level because atmospheric scintillation noise
appears to be a factor 3.6 smaller (Kenyon et al. \cite{photom})
than at the best temperate sites. The detailed issues for optical
photometry from Dome~C were highlighted by several authors, most
recently by Vernin et al. (\cite{vernin07}), Strassmeier
(\cite{kgs:roscoff}) and Kenyon et al. (\cite{photom}). We also note
the early attempts and experiences from the South Pole's
``Vulcan-South''
experiment\footnote{http:/\slash{}www.polartransits.org} that,
unfortunately, suffered from the comparable harsh winds at the South
Pole.

A continuous 1500-hour night opens up a new window for science cases
like the search for extra-solar planets (e.g. Pont \& Bouchy
\cite{pon:bou}, Deeg et al. \cite{deeg}), for asteroseismology (e.g.
Fossat \cite{foss:astro}) and for stellar rotation and activity
studies (e.g. Strassmeier \& Ol\'ah \cite{eddi}). Several
astronomical experiments are now planned for Dome~C (for a
compilation see Strassmeier et al. \cite{tenerife}). Among these
pilot experiments is a small 25-cm optical precursor telescope for
the 80-cm infrared telescope IRAIT (Tosti et al.~\cite{irait}) and
the 2$\times$60cm Schmidt-telescope ICE-T (Strassmeier et
al.~\cite{icet}). sIRAIT was designed to experience the expected
same difficulties in the same operative conditions foreseen for the
larger projects with the goal of performing a period of full tests
on a smaller scale.

In the current paper, we present and analyze the first data from
sIRAIT from one CCD field obtained in July during winter 2007. Our
primary target, V841~Cen, is a spotted, very active RS~CVn binary
with occasional flares and with an orbital period so close to a
multiple of a day (5.998 days) that time-series data from a
single, non-polar site severely undersamples its light curve. Our
secondary target is the $\delta$-Scuti star V1034~Cen, a
non-radially pulsating star in the lower part of the instability
strip with a (likely) fundamental period of $\approx$0.2 days.

\begin{figure*}[tbh]
\includegraphics[angle=-90,width=\textwidth,clip]{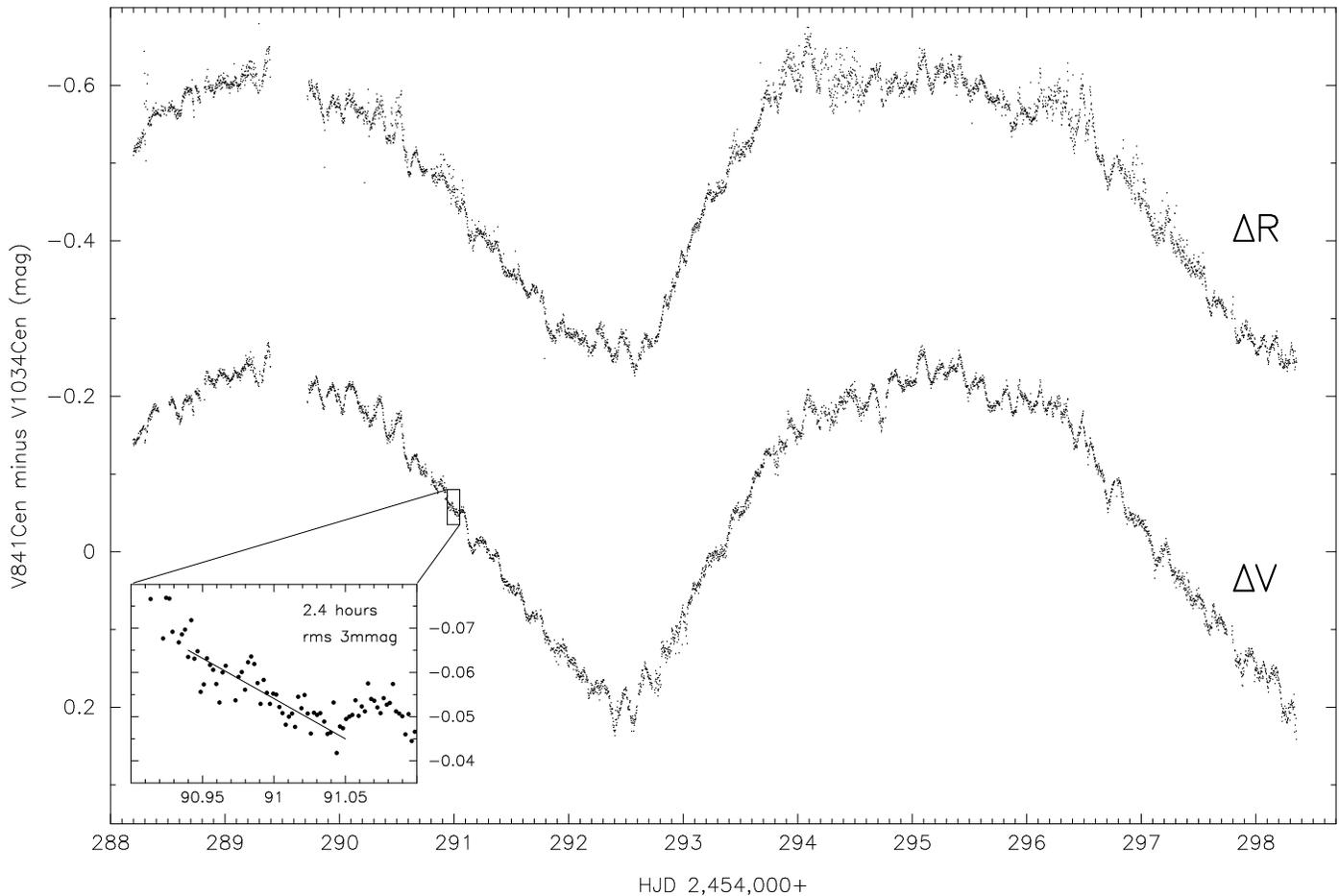}
\caption{Ten-day continuous differential $VR$ photometry of V841~Cen
minus V1034~Cen. Note that the long-period variation with an
amplitude of 0\fm4 in $V$ is due to spots rotating in and out of
view on V841~Cen while the short-period variations with an amplitude
of $\approx$0\fm02 are due to non-radial pulsations of V1034~Cen.
The insert shows a fraction of $V$ data that is nearly free of
intrinsic short-term stellar variations. The residuals from a simple
linear fit to a 2.4-hour subset suggest a rms scatter of just 3~mmag
in $V$ (4.2~mmag in $R$). For such a long duration, this is 3--4
times better than an equally sized telescope at a temperate site.
}\label{F2}
\end{figure*}

\section{Instrumental set up and observations}

\subsection{sIRAIT}

sIRAIT is a 25-cm, effective f/12 Cassegrain optical telescope on a
parallactic mount located near the \emph{Concordia} station in the
open field without a protection building (75$^{\circ}$06$'$04\arcsec
S, 123$^{\circ}$20$'$52\arcsec E, 3233m WGS84). Its CCD photometer
is located approximately 1m above the ground. The telescope was
designed by the IRAIT team at the University of Perugia, Italy
(Tosti et al. \cite{irait}) and built by the {\sl Marcon} telescope
factory of San Don\`{a} di Piave, Italy, equipped with electronics
and installed at Dome~C by one of the coauthors (R. Briguglio). It
is equipped with a guiding refractor placed along the optical tube
and moved by two extreme-environment stepper motors, originally
designed for vacuum and suitable for very low temperatures.

The focal-plane unit contains the CCD camera, the filter wheel, two
heaters, two fans, thermo-couples and Pt100 probes, the mirror
adjustment device and other electronics. It is insulated by a thin
layer of foam and it is internally heated by two resistance heaters.
The temperature achieved is constant to within $0.3\degr$C in the
inner CCD box, and $5\degr\pm2\degr$C in the motor-drivers box.
During acquisition the CCD temperature is set at $-28\degr$C. The
photometer is a commercial MaxCam CCD camera by Finger Lakes
Instruments (FLI). It is equipped with panchromatic Johnson $UBVRI$
filters and a focusing device. A Kodak KAF-0402ME CCD with
768$\times$512 9~$\mu$m pixels provides a field-of-view (FOV) of
8$\arcmin\times$5.3\arcmin\ with an image scale of
0.65\arcsec/pixel. Its quantum efficiency is given by the
manufacturer to 55\%\ in $B$, 80\%\ in $V$ and 60\%\ in $R$. The
full-well capacity is reached at 100,000 electrons and the FLI
controller allows a read-out-noise of 15 electrons at nominal
read-out speed of $\approx$500~kbit/s at a gain of 10\,$\mu$V/e$^-$.
From the flat-fields, we estimate a gain of 2.1e$^-$/ADU and a
read-out noise of 12.8e$^-$, according to the procedure summarized
in Janesick~(\cite{jan}). The typical point-spread-function (PSF)
measured from a $V$-band frame is a Gaussian with a full-width at
half maximum of 7.1 pixels (4.6\arcsec ) for a star with $V$=9\fm0,
30-sec exposure, 16,000 ADU at peak and S/N=12:1. Such an
over-sampled PSF minimizes many practical problems, e.g., the
contribution of partial pixels at the PSF's edge, the PSF changes
due to e.g. focus changes, tracking and guiding errors, wind shake,
or differential refraction but at the expense of increased crowding
by background stars.

\subsection{Acquisition and reduction of photometric data}

Both primary targets are in the same FOV of the CCD, itself
centered at $\alpha$ = 14$^h$34$^m$40$^s$ and $\delta$ =
--60$\degr$25\arcmin\ (2000.0) (Fig.~\ref{F1}). The telescope
acquired and tracked this FOV semi-automatically for a total of
243 continuous and consecutive hours (10.15 days) with only a
single 5.8-hour interruption (Fig.~\ref{F2}). Observations started
at HJD 2,454,288.199. A total of 13,000 CCD frames were acquired.
Note that the data were taken July 6-16, 2007, but the raw data
arrived in Europe only by the end of January 2008 when the
winter-over crew was able to leave the station.

\begin{table}[!tbh]
\begin{flushleft}
\caption{Log of three other variable stars in the CCD FOV. }
\label{T1}
\begin{tabular}{llllll}
\hline \noalign{\smallskip}
ID & NOMAD & $\alpha$ (in deg) & $V$        & $R$        & Notes \\
   &       & $\delta$ (in deg) & $\sigma_V$ & $\sigma_R$ & \\
\noalign{\smallskip} \hline \noalign{\smallskip}
3&\object{0295-0641140}&218.72680&9.57& 9.13& P=3.0d,1.5d\\
 &                     &-60.4277 &0.028& 0.028&\\
4&\object{0295-0640999}&218.70627&12.82&12.89&no clear P\\
 &                     &-60.4432 &0.051& 0.16&\\
7&\object{0296-0615795}&218.60366&13.14&13.17&no clear P\\
 &                     &-60.3841 &0.24& 0.25&\\
\hline
\end{tabular}
\end{flushleft}
\emph{Note:} ID is the internal identification according to
Fig.~\ref{F1}. ID=3 is the original comparison star CD-59\degr 5309.
NOMAD is the NOMAD catalog number (Zacharias et al. \cite{nomad}).
$\alpha$ and $\delta$ are for equinox 2000.0. $V$ and $R$ are their
average magnitudes in the Johnson system, and $\sigma_V$ and
$\sigma_R$ the standard deviations from the mean in the $V$ and $R$
bandpasses in magnitudes.
\end{table}

A sequence of observations consisted of consecutive $BVR$ frames
with 60s, 50s, and 40s integration times, respectively, typically
allowing for approximately 20 images per hour per filter with an
average time resolution of 155~sec. A daily $\approx$20--30 minutes
were needed for telescope derotation to unwind its electric cabling
but is barely noticeable in Fig.~\ref{F2}. A single 5.8-hour gap
occurred on the second day when the field was lost due to a tracking
error. The counts per pixel were on average half of the full-well
capacity of the CCD. The ratio of the sum of all background
corrected pixels within the aperture (approximately 20,000 pixels)
to the standard deviation of the background gives a peak S/N ratio
in $V$ and $R$ of 15,000:1 for V841~Cen ($V$=8\fm5). We did not
correct for cosmic-ray hits as these were rather seldom and did not
affect the photometry.

During the 10 days of acquisition the weather was very stable and
good, almost no wind, temperature around $-72\degr\pm2\degr$C.
Ground-layer seeing varied between 2.8\arcsec\ and an exceptional
5--6\arcsec. The few data from the latter periods had internal
deviations larger than 25~mmag and were rejected from the analysis
(a total of $\approx$20 frames out of 13,000). A simple linear fit
to a 2.4-hour long $V$ and $R$ data subset shows a standard
deviation of 3.0 and 4.2~mmag, respectively. We consider these as
upper limits because the selected data-subset may not be free of
intrinsic variability (see insert of Fig.~\ref{F2}).

\begin{figure}[tb]
    \centering
    \includegraphics[angle=-90,width=8.6cm]{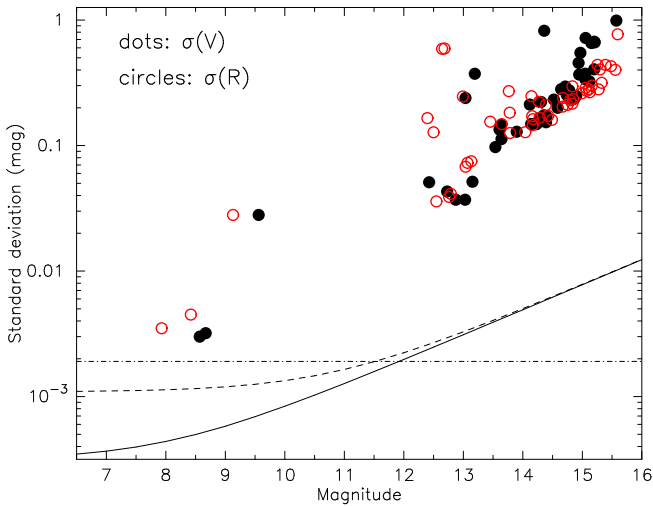}
    \caption{$V$ and $R$-band scatter diagram for an exposure time of 50~sec ($V$)
    and 40~sec ($R$). The precision for the two target stars ($V\approx$8\fm5)
    was 3~mmag in $V$ and 4.2~mmag in $R$. The lines indicate the
    sum of the photon and scintillation noise for a temperate site
    (dashed line) and for Dome~C (full line). The horizontal dash-dotted line is
    an \emph{estimate} of the convolved sky plus detector limit for sIRAIT.}
    \label{F3}
\end{figure}

Calibration frames were obtained three days after the science
observations and consisted of 20 twilight flat fields in $BVRI$ and
series of biases. Flat fields were acquired from horizon pointing
during ``midday'' when the Sun was still below the horizon but
provided enough light for exposures. Exposures of 30s, 20s, and 10s
for $BVR$, respectively resulted on average in 20,000 counts. A
data-cube fit to all 20 flat-field frames was performed to obtain a
master flat. Dark frames were taken occasionally during the
cable-derotation times and sum up to 18, 25, and 26 frames for $B$,
$V$, and $R$, respectively.

Twenty-five standard stars from Landolt (\cite{land}; and references
therein) were observed ten times each on a clear night on September
10th, 2007 at the end of the Antarctic night. The air mass range was
between 2--3 and thus unfortunately always higher than for our
science field. Large air-mass variations are impossible to obtain at
Dome~C because of its high geographic latitude. Images were acquired
in $UBVRI$ but $U$ suffered from very poor S/N and was discarded.
Even $B$ had comparably large scatter. Extinction coefficients of
0\fm278, 0\fm179, 0\fm064, and 0\fm081 for $B$, $V$, $R$ and $I$,
respectively, were obtained from linear fits to the instrumental
magnitudes versus air mass and then used to transform to the
standard Johnson system (Briguglio \cite{runa}).

The designated comparison star was \object{CD-59\degr 5309}.
\emph{Simbad}\footnote{http://simbad.u-strasbg.fr/simbad/} lists it
as a B-star with $V$=9\fm50, $B-V$=+0\fm74 and $U-B$=$-$0\fm27 as
its best-quality $UBV$ values from Schild et al. (\cite{schild}).
Our data show it to be a low-amplitude variable star with
$V$=9\fm57, a 1$\sigma$ scatter of 0\fm028, and $V-R$=0\fm44. To
increase its S/N ratio, we summed up once three and once ten
consecutive $V$ frames and then performed a period analysis on them.
However, the noise in the summed 10-frame light curve appeared
sometimes comparable to the individual frames. We attribute this to
the jitter from tracking errors which sometimes even includes jumps.
Nevertheless, both data sets revealed periods of 3.04 days and 1.54
days as the strongest peaks (both in a Scargle and a CLEAN
application; see Sect.~\ref{S41}) but with rather low amplitudes
comparable to the scatter of the data. Two more periods of 5.57 and
0.983 days appear significant but have even lower amplitudes while 6
more periods are possibly in the data but are judged unreliable due
to the large amplitude of the noise. The residuals from a
least-squares fit with all four periods is 0.05 mag. We conclude
that the star is likely a non-radially pulsating B star with a
fundamental period of either 3.0 or 1.5 days.

The CCD FOV contains another 60 stars that we were able to identify
in NOMAD (Zacharias et al. \cite{nomad}) with $V$ magnitudes between
12\fm4--15\fm7. None of these are identified in \emph{Simbad}
though. Two stars are possibly variable objects due to their
higher-than-expected standard deviation and are listed in
Table~\ref{T1}. No clear periods were found for either.

We employed the ARCO software package for CCD data reduction and
analysis as described in Distefano et al.~(\cite{arco}). Standard
CCD frame reduction consisted of bias and dark subtraction and
flat-field division. We use the SExtractor (Bertin \& Arnouts
\cite{sex}) software to match stars detected in different CCD
frames, select the best candidates for building an ensemble of
comparison stars and identify the parameters that optimize the
photometry of each star and minimize the scatter of light curves due
to statistical fluctuations. Aperture photometry with an optimized
aperture is then applied to all selected candidates between
$R$=7\fm9 and 15\fm7. Fig.~\ref{F3} shows the scatter plot for $V$
and $R$ from a three-comparison star ensemble solution. We note that
some stars were located so close to the edge of the FOV that they
were sometimes not identified by the photometry package which then
resulted in artificially high residuals. These stars are not plotted
in Fig.~\ref{F3}. An eye-ball fit to the lower envelope from
$\approx$50 stars suggests a photometric precision of sIRAIT for the
range 12\fm5 to 15\fm5 of 0\fm04 at 12\fm5 and 0\fm4 at 15\fm5. For
the very bright targets the precision is approximately 2.5~mmag
above the expected scintillation-noise limit of 0.5~mmag (the full
line in Fig.~\ref{F3}) while the faint stars are basically
instrumental-noise limited (for a detailed discussion see Newberry
\cite{new}).

\begin{figure}[tb]
    \centering
    \includegraphics[angle=-90,width=8.6cm]{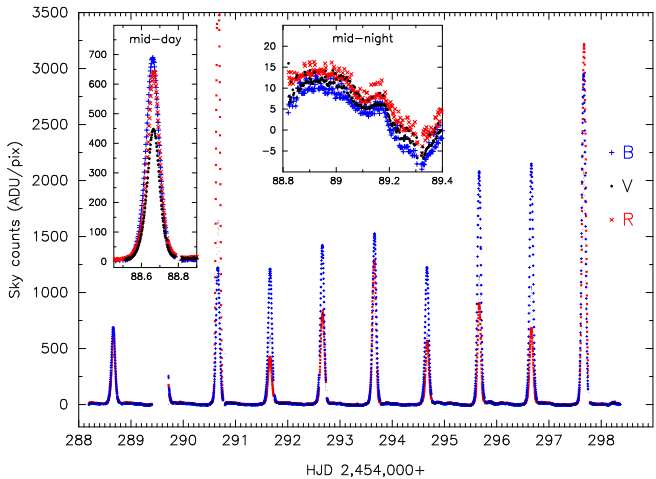}
    \caption{Average background counts in $BVR$ for the ten days
    of our observations. The inserts show an example of the counts
    during midday (left) and midnight (right), respectively. $BVR$ is indicated with
    different plot symbols (pluses $B$, dots $V$, crosses $R$).}
    \label{F4}
\end{figure}

An estimate of the Dome-C sky brightness is obtained by using the
background counts given by SExtractor on the dark-subtracted input
frames (Fig.~\ref{F4}). We discarded all frames affected by
twilight, i.e. during fractional JD of 0.5 to 0.85 as seen in
Fig.~\ref{F4}. Note that moonlight did not contaminate our frames as
we had waning moon. The maximum height of the moon above the horizon
was 6\degr\ during the first night and 4\degr\ during the second
night, after that it was continuously below the horizon reaching new
moon two days before the end of our campaign. The remaining frames
showed average background count rates of 2.99, 4.65, and 6.83
ADU/pix in $B$, $V$, and $R$, respectively, at a rms of around
6.5~ADU. Converting these to sky brightness per square-arcseconds
yields magnitudes of 20\fm77, 20\fm36, 19\fm90 in $B$, $V$, and $R$,
respectively. We estimate that these values are good to within only
$\approx$0\fm1. Examples are shown in the inserts of Fig.~\ref{F4}.
Our $V$ value is possibly significantly brighter than the
GATTINI-SBC estimate of $\approx$21 Vmag/arcsec$^2$ converted from a
Sloan $g$ filter (Moore et al.~\cite{moore}). Surprisingly, the
background counts at midday varied by as much as a factor of four in
$V$ and $B$, and up to a factor of seven in $R$ which may have
resulted from illuminated high clouds. The ten-day rms at midnight
was rather stable though and converts to a sky plus detector limit
of 1.88\,mmag in $V$. This is shown in Fig.~\ref{F3} as a straight
line.

\subsection{Problems encountered}\label{S23}

A spatially non-uniform gain of the CCD, dependent also on
count-rate and time, did not allow the use of the original
comparison star for high-precision differential photometry. This
behavior was not noticed in pre-shipment CCD tests and is likely due
to the controller or the environment in which it has been working in
Antarctica. However, the $B$ bandpass was irreparably affected by
this due to its already low count rates and we did not use it for
further analysis.

Photometry with the full ensemble solution with 20 comparison stars
resulted in more than twice the scatter for the two main targets
than with respect to the differential magnitudes of the two bright
stars themselves. The selection of the three next brightest and
closest stars to V841~Cen as comparison stars resulted in a better
but still more scattered light curve. Notice that all these
bona-fide comparison stars are at least four magnitudes fainter than
our two target stars, and the differential magnitudes are
accordingly noisier. We bypassed this, and the gain problem, by
using the $\delta$-Scuti star V1034~Cen as the comparison star for
V841~Cen (and vice versa) because it is the only star in the FOV
with almost identical count rates.

Another problem occurred when the optics were cleaned twice during
the night (at HJD 2,454,292.68 and 93.71). Although technically not
related with it, it likely caused another CCD controller (or USB
connection) problem due to unwanted interference. On both occasions
it had suddenly decreased the count rate for V841~Cen, but increased
the count rate for V1034~Cen (on the same exposure). Both count
rates then exponentially resettled to the previous (expected) count
rates within $\approx$10 hours. We have no ready explanation for
this but approximated the resettling trend with an exponential fit
to the (differential) data and removed it. This introduced an extra
scatter of $\approx$1~mmag during these 10 hours so that the
standard deviation during these hours was more like
$\sqrt{3^2+1^2}\approx 3.2$~mmag.

Next, we noticed that the telescope had severe pointing and tracking
errors that accumulated during the run. Repositioning was done
manually after approximately an hour or so. A scatter plot of the
central coordinates of all, e.g., $V$ frames shows an elongated
distribution with a peak-to-peak range of 100\arcsec\ in both
declination and right ascension. This means that the FOV usable for
continuous photometry is just the minimum wrapped-in field from all
individual pointings. With a CCD FOV of 8$\arcmin\times$5.3\arcmin\
the range of $\pm$50\arcsec\ is a significant restriction. This
probably impacted on the photometric precision because the
photometric aperture did not always enclose exactly the same pixels.

\begin{figure}[tb]
\includegraphics[angle=0,width=8.6cm,clip]{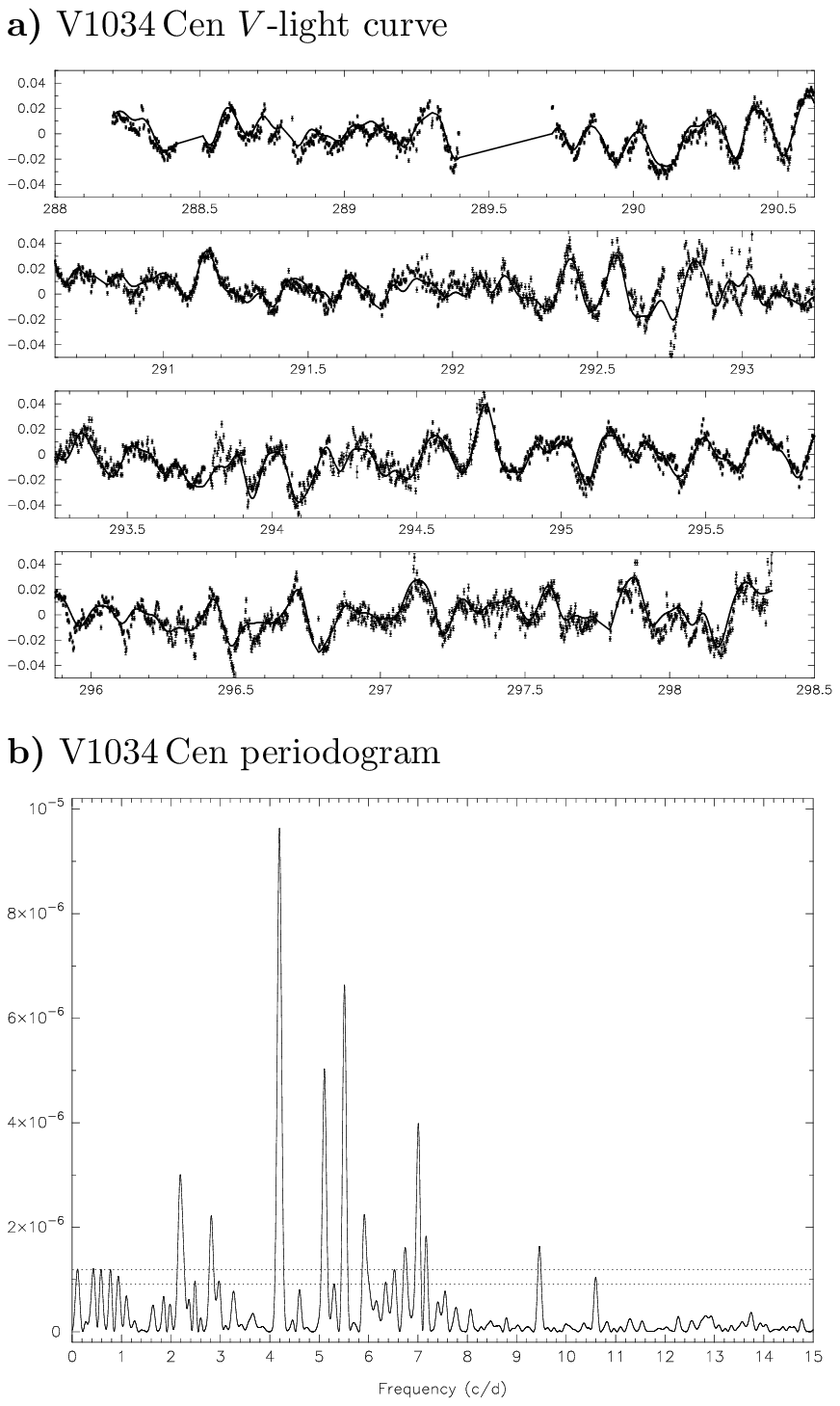}
\caption{{\bf a} $V$-light curve of V1034~Cen after reconstructing
and subtracting the variation of V841~Cen. Time is in fractional
Julian Date as in Fig.~\ref{F2}. Error bars are $\pm 1\sigma$. The
line is the least-squares fit with the frequencies in
Table~\ref{T2}. {\bf b} Period determination from a rectified
CLEAN periodogram. The two horizontal lines indicate a false alarm
probability of 10$^{-6}$ (lower line) and 10$^{-9}$ (upper line),
respectively. A total of 20 periods between 2.2 hours and 3.5 days
appear above the 10$^{-6}$ FAP. Note the complete absence of the
one-day period and its aliases. }\label{F5}
\end{figure}

\section{The primary target stars}

\subsection{V841 Cen = HD127535}

\object{V841 Cen} ($\alpha = 14^h34^m16^s$, $\delta =
-60^\circ24\arcmin27\arcsec$, 2000.0, $V$=8\fm5) is a rapidly
rotating, single-lined spectroscopic binary with an active K1
subgiant as the primary component (Collier \cite{acc82a}). The star
exhibits strong Ca\,{\sc ii} H\&K and H$\alpha$ emission (Houk \&
Cowley \cite{hou:cow}, Weiler \& Stencel \cite{wei:ste}). It shows
high X-ray flux in the ROSAT 0.1--2.4 keV energy range (Dempsey et
al. \cite{demp}) and in the EUV (Mitrou et al.~\cite{euv}), and also
very high radio-flux densities (Slee \& Stewart \cite{sle:ste}). Its
lithium abundance of $\log n$=0.77 (Barrado y Navascu\'es et
al.~\cite{barrado}, but see also Randich et al.~\cite{rand} who
obtained a significantly higher value) suggests a comparably young
system. Randich et al. (\cite{rand}) determined a $v\sin i$ of
33$\pm$2 km\,s$^{-1}$ while De\,Medeiros et al. (\cite{dem:don})
obtained 10$\pm$1 \kms\ from CORAVEL tracings.

The orbit is circular with a period of 5.998 days (Collier 1982a),
while the photometric (= rotational) period of the K1 subgiant was
given by Cutispoto (\cite{c90}) to be 5.929$\pm$0.024 days. Thus,
the orbital motion and the stellar rotation are bound but not
quite exactly synchronous and/or the subgiant's surface is
differentially rotating.

Previous photometry of V841~Cen was presented by Collier
(\cite{acc82b}) from 1980--81,  by Innis et al. (\cite{inn:coa})
from 1981, by Udalski \& Geyer (\cite{uda:gey}) from 1984, by Bopp
et al. (\cite{bopp}) from 1985, by Mekkaden \& Geyer
(\cite{mek:gey}) from April 1987, by Cutispoto (\cite{c90}) from
February 1987, by Cutispoto (\cite{c93}) from 1989, by Cutispoto
(\cite{c96}) from Feb.-March 1990, by Cutispoto (\cite{c98a}) from
March 1991, by Cutispoto (\cite{c98b}) from February 1992, by
Strassmeier et al. (\cite{ibvs}) from June--July 1994 and, most
recently, by Innis \& Coates (\cite{innis}). At one occasion,
Cutispoto (\cite{c98a}) detected a flare lasting at least one day.
V841~Cen is listed as star number CABS\,118 in the ``Catalog of
Chromospherically Active Binary Stars'' (Strassmeier et al.
\cite{cabs}).

\subsection{V1034 Cen = HD127695}

\object{V1034 Cen} ($\alpha = 14^h35^m01^s$, $\delta =
-60^\circ23\arcmin32\arcsec$, 2000.0, $V$=8\fm7) is an A9IV
$\delta$-Sct star with a period of 0.235 days and a full amplitude
of 0\fm03 in $V$ (Koen et al. \cite{koen}). Only one period is
known, so it is likely the fundamental period. A summary of known
parameters was given by Rodriguez et al. (\cite{rod}) in the
revised $\delta$-Sct catalog. The photometry by Koen et al. was
obtained during 18 hours on two consecutive nights and analyzed
together. The authors mentioned that the $V$- and $B$-band phases
were statistically identical but hinted that there may be
additional low-amplitude frequencies. Koen et al. (\cite{koen})
pointed out some target confusion in the Geneva photometry catalog
(Rufener \cite{rufener}) that likely came from observing V841~Cen
instead of V1034~Cen.

\begin{figure*}[tbh]
\includegraphics[angle=0,width=\textwidth,clip]{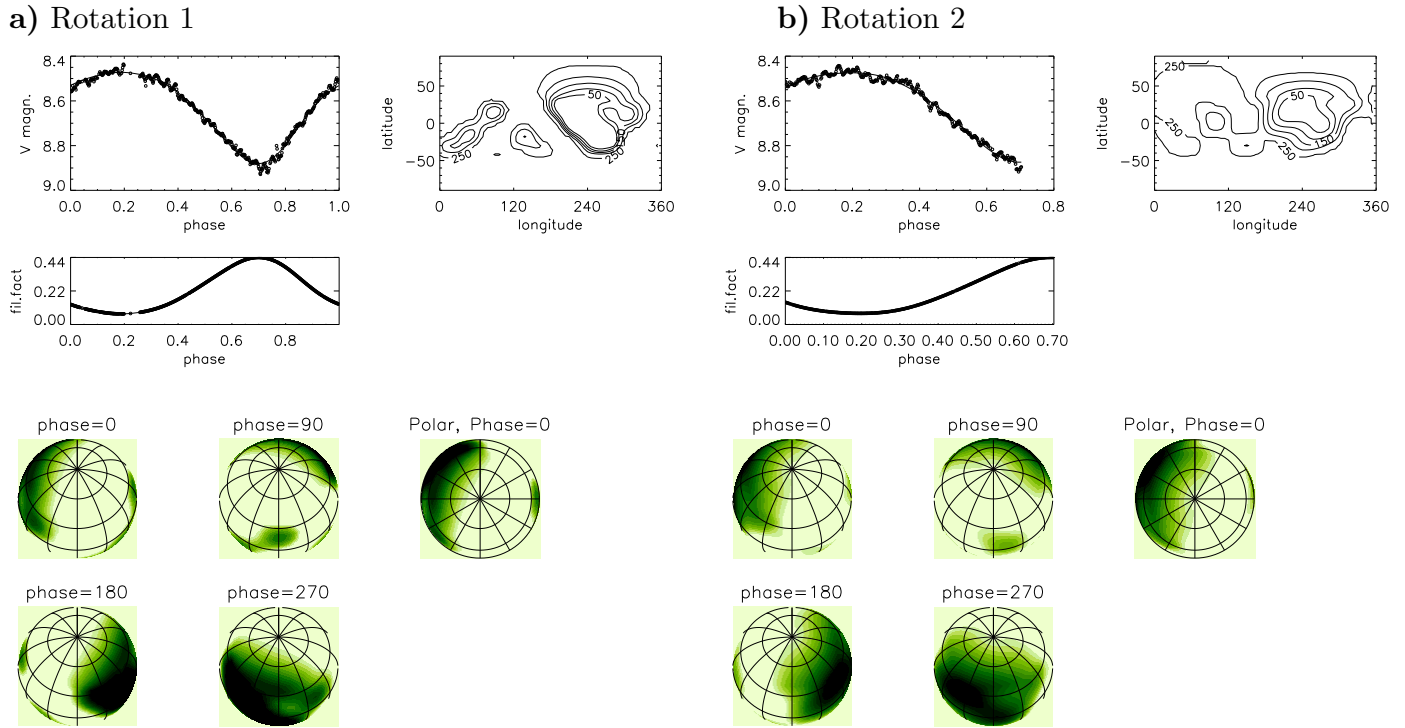}
\caption{A spot model for V841~Cen. {\bf a} Stellar rotation \#1 ,
{\bf b} Stellar rotation \#2. Each panel shows the fit to the
$V$-band data (top left), the spot-filling factor in \% per
hemisphere as a function of rotational phase (middle), a contour
spot map in Mercator projection (top right), and maps of the stellar
surface in spherical and flattened pole-on view (entitled polar) in
the lower panels. Note that phase zero coincides with our first data
point. }\label{F6}
\end{figure*}

\begin{table}[tb]
\begin{flushleft}
\caption{Period identifications for the $\delta$\,Sct-star
V1034~Cen. }\label{T2}
\begin{tabular}{lllll}
\hline \noalign{\smallskip}
No & Frequency & Period & Amplitude & Notes \\
   & (cycle/d)    & (days) & (mmag) &  \\
\noalign{\smallskip}\hline \noalign{\smallskip}
1  & 4.200 &0.238 & 12.04 &fundamental\\
2  & 5.514 &0.181 &  9.91 &stable\\
3  & 0.145 &6.904 &  9.75 &uncertain\\
4  & 0.461 &2.169 &  9.71 &uncertain\\
5  & 2.485 &0.402 &  9.33 &\\
6  & 5.107 &0.196 &  8.90 &stable\\
7  & 7.011 &0.143 &  8.14 &stable\\
8  & 0.766 &1.304 &  7.80 &uncertain\\
9  & 0.286 &3.492 &  7.35 &uncertain\\
10 & 2.192 &0.456 &  6.87 &\\
11 & 5.909 &0.169 &  6.20 &stable\\
12 & 6.738 &0.148 &  5.94 &stable\\
13 & 2.404 &0.416 &  5.80 &\\
14 & 7.161 &0.140 &  5.78 &stable\\
15 & 2.552 &0.392 &  5.68 &\\
16 & 9.462 &0.106 &  5.09 &stable\\
17 & 2.795 &0.358 &  5.04 &\\
18 & 2.941 &0.340 &  5.02 &\\
19 & 4.167 &0.240 &  4.92 &\\
20 & 6.509 &0.154 &  4.41 &\\
21 & 6.353 &0.157 &  4.37 &\\
22 &10.600 &0.094 &  4.11 &stable\\
23 & 6.026 &0.166 &  4.10 &\\
24 & 5.294 &0.189 &  4.04 &\\
\hline
\end{tabular}
\end{flushleft}
\emph{Note:} The amplitude is the peak-to-valley $V$ amplitude in
milli-mag. Frequencies no.~11-24 are formally below the 1$\sigma$
quality of the fit but are identified in the periodogram analysis in
Fig.~\ref{F5}b. A ``Stable'' flag means that a frequency is mostly
independent of other frequencies, ``uncertain'' means that the
frequency is likely influenced by the length of the data set.
\end{table}

\section{Results and discussion}

Fig.~\ref{F2} shows the entire time series V841~Cen minus
V1034~Cen for Johnson $V$ and $R$. Note again that the comparison
star CD-59\degr 5309, one-magnitude fainter than the two target
stars, could not be used as such due to a CCD-controller problem.
Therefore, our first step was to separate the light variability of
the two stars. Fortunately, this can be done quite accurately
because of the star's significantly different variability periods
and amplitudes.

\subsection{The rotation period of V841\,Cen}\label{S41}

Our periodogram analysis for the combined V841~Cen minus V1034~Cen
$V$ data prominently shows just the expected single frequency of
around six days. We independently apply four different period-search
routines to the combined differential data and then examine the
pre-whitened output. Phase Dispersion Minimization (PDM, Lafler \&
Kinman~\cite{Laf:Kin}, Stellingwerf~\cite{Stel}) and Lomb-Scargle
(LS, Scargle \cite{ls}) result in a comparably broad $\chi^2$
minimum and consequently a just moderately-well determined period of
5.881 days and 5.884 days, respectively. The Minimum String Length
(MSL, Dworetsky \cite{msl}) and the CLEAN algorithm (Roberts et al.
\cite{clean}) provide a significantly sharper minimum and also give
marginally longer periods of 5.8872 days and 5.8854 days,
respectively. Just because of consistency reasons we adopt the CLEAN
period as the most likely rotation period of V841~Cen with an
accuracy based on the rms of the four periods (0.0026 days). Note
that an internal error of 10$^{-6}$ days is obtained from the width
of its $\chi^2$ minimum via the criterium of Bevington (\cite{bev}).

The surface rotation is thus synchronized to within 2\% of the
orbital period and suggestive of an older system, which appears to
be somewhat in contradiction with its relatively high lithium
abundance. The K subgiant has an upper limit Li abundance that is
about a factor 10 below what is considered to be a lithium-rich
star but appears with a higher than normal Li surface abundance
(do Nascimento et al.~\cite{donas}). High degrees of
synchronization of the magnetically active component in RS~CVn
binaries is typical of the vast majority of systems with orbital
periods up to 30 days (Fekel \& Eitter \cite{fe89}).

\subsection{The pulsation spectrum of V1034\,Cen}

The base rotational frequency of V841~Cen and up to 6 of its
higher-order multiples are identified and subtracted from the
combined light curve. The remaining V1034~Cen contribution is shown
in Fig.~\ref{F5}a along with the least-squares fit from a total of
41 frequencies of which 24 are ranked significant according to the
criterium of Breger (\cite{breger}), which suggests a 99.9\%
probability for a peak not to be generated by noise if the obtained
amplitude S/N exceeds 4.0. Note that the least-squares fit was
obtained from the combined light curve, including the 7 frequencies
needed to describe the V841~Cen light curve, but only the
pre-whitened output is plotted in Fig.~\ref{F5}a. The final $\chi^2$
achieved was 6.8~mmag, close to the average photometric precision.
The periodogram from just the reconstructed V1034-Cen data is shown
in Fig.~\ref{F5}b. Using the CLEAN approach, a total of 10 periods
appear above a false-alarm probability (FAP) of 10$^{-9}$ with
peak-to-valley amplitudes in the range up to 12~mmag. Still 10 more
periods appear above a FAP of 10$^{-6}$ with amplitudes below
$\approx$6~mmag. Our frequency resolution from the full width at
half maximum of the spectral window is 0.062 cycle/d. Frequencies
shorter than $\approx$0.2~cycle/d are problematic because these get
close to the rotation cycle of the spotted star (0.17 cycle/d) and
could be misinterpreted. E.g., a frequency of 0.145 cycle/d
(Table~\ref{T2}) has formally an amplitude of 9.75~mmag and would be
highly significant but is judged particularly uncertain because its
cycle length of 6.9 days is close to the length of the entire data
set and close to the rotation period of V841~Cen. In Table~\ref{T2}
we list all frequencies that we consider significant with respect to
the data quality and the periodogram analysis. Note that the
original frequency of $\approx$4.2 cycle/day (a period of
$\approx$0.2 days) found by Koen et al. (\cite{koen}) is
reconstructed as the strongest peak in our data set in
Fig.~\ref{F5}b at 4.20018 cycle/d, which nicely confirms Koen et
al.'s initial discovery.

\subsection{A spot model for V841 Cen}

We employ the new light-curve inversion code of Savanov \&
Strassmeier (\cite{sav:str}). It reconstructs the stellar surface
spot configuration from multi-color light curves by using a
truncated least-squares estimation of the inverse problem's
objects principal components. Our unknown object, the photospheric
spot-filling factor, is a composite of a two temperature
contribution; the intensity from the photosphere, $I_{\rm P}$, and
from cool spots, $I_{\rm S}$, weighted by the fraction $f$ of the
surface covered with spots, i.e. the spot filling factor. The
intensity per pixel reads then
\begin{equation}\label{f:1a}
I = f \times I_{\rm P} + (1 - f) \times I_{\rm S}
\end{equation}
with $0< f <1$.  The inversion results in the distribution of the
spot filling factor over the visible stellar surface that best
fits the data. No assumptions for the shape or for the
configuration or total number of spots are made. The stellar
astrophysical input includes band-pass fluxes calculated from
atmospheric models from Kurucz (\cite{K00}).

An effective temperature for V841~Cen of 4390~K was listed in
Barrado y Navascu\'es et al.~(\cite{barrado}) who based it on $V-I$
and $R-I$ indices. Note that just $B-V$ alone would suggest a 300-K
higher temperature of $\approx$4,700\,K based on, e.g. Flower
(\cite{f96}). Karatas et al.~(\cite{kar}) gave a spectroscopic
parallax that places V841~Cen at a distance of 63$^{+26}_{-13}$ pc
and thus with an accordingly uncertain luminosity of $\approx$2.3
L$_\odot$ (no \emph{Hipparcos} parallax is available) . Our best
rotation period of 5.8854~days and the projected rotational velocity
of 10$\pm$1 \kms\ from De\,Medeiros et al. (\cite{dem:don})
determine together the lower limit of the stellar radius to $R\sin
i$=1.16$\pm$0.12~R$_\odot$. Assuming the luminosity based on the
spectroscopic parallax and the Stefan-Boltzmann law we obtain a
matching radius only with inclinations of as low as 30\degr\ and
26\degr\ for effective temperatures of 4700~K and 4400~K,
respectively. Cutispoto (\cite{c98b}) favored a K3(V-IV) spectral
type from the observed long-term $UBVRI$ colors which leads to a
slightly higher inclination of $\approx$40\degr\ ($\pm$10\degr ) and
$T_{\rm eff}$=4500~K. These are the values we adopt for the spot
modelling. A generally low inclination is also in agreement with the
low mass function of $f(m)$=0.025 from the orbit by Collier
(\cite{acc82a}) and the fact that we do not see eclipses nor a
secondary star in any of the published spectra (e.g. Randich et al.
\cite{rand}). In any case, the numerical light-curve simulations by
Savanov \& Strassmeier (\cite{sav:str}) showed that a change of the
inclination angle of even $\pm$15\degr\ just marginally altered the
light-curve solution.

Due to the continuous time coverage we do not need to convert the
data into phase space but separate into first and second rotation
based on a period of 5.8854 days. Five consecutive data points were
always averaged. Fig.~\ref{F6} shows the results for the consecutive
1.7 stellar rotations, dubbed ``rotation 1'' and ``rotation 2''. A
huge spot covering up to 44\% of the visible hemisphere is required
to fit the deep 0\fm4 photometric minimum. A second, smaller spot
with $\approx$10\% filling factor at a longitude of
$\approx$100\degr (phase 0\fp28) located in the adjacent hemisphere
is needed to fit the broad shape of the light curve near maximum
light. Note that our inversion algorithm converges with a $\chi^2$
of the fit that is always the (average) $\chi^2$ of the data.
Despite the low inclination of the rotational axis, and the
therefore pronounced pole-on view, the inversion reconstructs both
spots at low latitudes rather than at the poles. This is due to the
deep and relatively sharp photometric minimum that excludes a polar,
permanently-in-view location.

A spot coverage of 44\% of the visible hemisphere is among the
largest measured values for active stars, and is by chance the same
value as determined for the largest spot ever recorded with the
Doppler-Imaging technique (for XX~Tri, a K-giant in a 24-day RS~CVn
binary; Strassmeier \cite{xxtri}). However, spot sizes obtained from
photometry depend on the spot temperature. The full $\Delta(V-R)$
amplitude in our data is 0\fm090$\pm$0.004, becoming redder during
minimum brightness and bluer during maximum brightness. Our
inversion code resolves this with a most likely spot temperature of
3750~K. Its error is obtained from the numerical simulations in
Savanov \& Strassmeier (\cite{sav:str}) that suggested that $f$
increases up to 30\% if $\Delta T=T_{\rm P}-T_{\rm S}$ is lowered by
250~K. We conclude that $f$=44$\pm$3\% and $\Delta T$=750$\pm$100~K
are the most likely spot parameters for V841~Cen at the time of our
observations.

Because the orbit determination is almost 30 years old and had a
just modestly precise period, it is impossible to identify the exact
orbital phase for the time of our data in 2007. Therefore, no
statement can be made regarding the location of the largest spot
with respect to the orbital frame. However, this would be needed to
interpret the magnetic-flux emergence in such a binary because the
proximity of the companion star breaks the rotational symmetry and
cause a non-uniform surface flux distribution (e.g.
Holzwarth~\cite{holz}).

\section{Conclusions and outlook}

We presented 243 continuous hours of optical photometry from
Antarctica with a duty cycle of 98\% and a cadence of 155 seconds. A
3~mmag rms precision in $V$ over 2.4 hours with the 25-cm sIRAIT
telescope was achieved for the two bright 8\fm5 target stars. This
is a factor 3--4 better than what we had obtained with the 25-cm T1
automatic photoelectric telescope (APT) at Fairborn Observatory in
southern Arizona (Strassmeier et al.~\cite{str:bar}, Henry
\cite{h95a}). Most likely this is attributed to scintillation noise
smaller by a factor 3--4 compared to temperate observing sites, as
reported by Kenyon et al. (\cite{photom}). We conclude that
high-precision continuous photometry within the turbulent ground
layer just one meter above ground is feasible at Dome~C, even with
low-cost, partly commercial components. The main problems we had
encountered with sIRAIT in its first winterover were all due to the
quality of the equipment rather than with the harsh Antarctic
environment. This makes us strongly believe that our proposed
2$\times$60-cm, more optimized and robust, photometric facility
ICE-T (Strassmeier et al.~\cite{icet}) as well as the
intermediate-step 40cm \emph{a-step} experiment (Fressin et al.
\cite{astep}) are well suited for a site like Dome~C and could, for
some favorable cases, even challenge photometry from space.

\acknowledgements{The field activities and the results at Dome~C
benefit from the support of the French and Italian polar agencies
IPEV and PNRA in the framework of the Concordia station programme.
We tank the AstroConcordia astronomers D.~M\'ekarnia and
F.~Jeanneaux for their support during the winterover. Heidi Korhonen
kindly provided the scintillation noise values for temperate sites.
We thank an anonymous referee for several helpful suggestions that
improved the paper. German participation in sIRAIT was financed by
AIP through the State of Brandenburg and the federal Ministry of
Education and Research and supported by the German polar agency AWI.
Discussions with Michel Breger on the significance of pulsation
frequencies is also appreciated. Finally, we acknowledge support
from the European Community's Sixth Framework Programme under
contract number RICA-026150 (ARENA). }

\end{document}